\documentclass[12pt]{article}
\usepackage{amssymb,amsmath,graphicx}
\begin{document}
\title{\textbf{Counting Gribov copies}} 
\author{B.~Holdom%
\thanks{bob.holdom@utoronto.ca}\\
\emph{\small Department of Physics, University of Toronto}\\[-1ex]
\emph{\small Toronto ON Canada M5S1A7}}
\date{}
\maketitle
\begin{abstract}
We show that an explicit counting of Gribov copies can shed light on the infrared behavior of non-abelian gauge theories. A power-law growth of the number of copies suppresses gluon propagation while the distribution of copies along a gauge orbit implies an enhanced density of small eigenvalues of the Fadeev-Popov operator. Both of these phenomena are related to confinement. The discreteness in the number of copies and the associated nonlocality also has implications for vacuum energy.
\end{abstract}

\section{Introduction}
The continuum definition of non-abelian gauge theories is sensitive to the global structure of the field configuration space, since gauge equivalent configurations remain after standard gauge fixing procedures are applied \cite{n1}. Rather than just a technical nuisance, the structure of configuration space that is implied by the existence of Gribov copies may be central to the understanding of a mass gap and the mechanism of confinement in these theories.

A gauge fixing condition $G(A)=0$ defines a hypersurface in the gauge field configuration space. But a gauge orbit, a set of gauge equivalent configurations, can intersect this hypersurface more than once. To characterize this we consider the following integration along a gauge orbit where $A^U_\mu=U^\dagger A_\mu U+\frac{1}{g}U^\dagger\partial_\mu U$,
\begin{equation}
1+N(A)=\int [dU]\delta(G(A^U))\left|\det\left(\frac{\delta G(A^U)}{\delta U}\right)\right|.
\label{e13}\end{equation}
This result has an analog in ordinary calculus. $\delta G(A^U)/\delta U$ is the Fadeev-Popov (FP) operator which we henceforth denote by ${\cal D}_A$.  $N(A)$ is the number of Gribov copies, the number of additional intersections of the orbit with the hypersurface, and it is clear that $N(A^U)=N(A)$.

One source of Gribov copies arises from the existence of different regions of configuration space in which the spectrum of ${\cal D}_A$ has different numbers of negative modes. At a boundary between two such regions an eigenvalue of ${\cal D}_A$ is changing sign. Gribov proved that pairs of copies can exist close to boundaries, with one copy on each side. The first Gribov region ${\cal D}_A$ has no negative eigenvalues, and this region includes the perturbative configurations. Gribov copies can be ignored in the perturbative limit.

The generating functional for the quantum theory can account for the presence of Gribov copies as follows \cite{n1},
\begin{equation}
{\cal Z}[J]=\int [dA] e^{iS[A,J]}\delta(G(A))\left|\det({\cal D}_A)\right|\frac{1}{1+N(A)}.
\label{e1}\end{equation}
$S[A,J]$ is the standard pure gauge theory action with source $J$. Given that ``this problem seems to be almost hopeless'' \cite{n1} given the presence of $N(A)$, attention was focused on attempts to directly excise copies from the functional integral. Of particular interest is the fundamental modular region (FMR), which is a subregion of the $G(A)=0$ hypersurface such that every gauge orbit intersects it once and only once. It is contained within the first Gribov region although there is some overlap of the boundaries of the two regions \cite{a13}.

From the definition of the FMR we can write $[dA]=[dA^{\rm FMR}][dU]$ (see \cite{n13} for more details), and then from (\ref{e13}) we have
\begin{equation}
{\cal Z}[J]=\int [dA^{\rm FMR}] e^{iS[A,J]}.
\end{equation}
This restriction on the functional integration is difficult to implement in a continuum description due to the highly nontrivial nature of the FMR. Instead a restriction to the larger first Gribov region is often considered since it can be implemented as a positive definite constraint on $\det({\cal D}_A)$. Then the absolute value in (\ref{e1}) may be removed and it is also argued that remaining copies are such that the $1/(1+N)$ factor can be neglected \cite{n12}.

Our approach in this work will be to remain true to (\ref{e1}) and allow the functional integral to sample the Gribov copies. A prototype of this type of approach to Gribov copies was studied in detail in \cite{n6}. One goal then is to obtain information about $N(A)$ directly through explicitly counting copies. In addition we shall argue that the distribution of copies along a gauge orbit provides information about the spectrum of ${\cal D}_A$. The behaviors of $N(A)$ and ${\cal D}_A$ that we find have implications for the confinement mechanism.

The Landau gauge condition $G(A)=\partial^\mu A_\mu=0$ gives ${\cal D}_A=-\partial^\mu D_\mu$ with $D_\mu$ the standard gauge covariant derivative while the Coulomb gauge condition $G(A)=\vec{\nabla}\cdot \vec{A}=0$ gives ${\cal D}_A=-\vec{\nabla}\cdot \vec{D}$. Gribov copies exist in both gauges. Coulomb gauge lacks Lorentz invariance and is in fact is not a complete gauge fixing even without Gribov copies. But it provides a more direct connection to the confinement mechanism through the long distance behavior of the instantaneous color Coulomb potential, where the latter is governed by the low momentum spectrum of ${\cal D}_A$. Also, since ghosts are absent in the Coulomb gauge,\footnote{The cancellation of $|\det({\cal D}_A)|$ due to the $A_0$ integration is described for example in \cite{n10}.} we can infer that the ghosts present in the Landau gauge will also be absent from the physical states. This must be the case even though (\ref{e1}) lacks a Becchi-Rouet-Stora-Tyutin (BRST) symmetry.  In fact it is presently not known how to maintain a local BRST invariance while accounting for Gribov copies \cite{n13,a21}. BRST symmetry emerges in the perturbative limit where Gribov copies are absent, and then the construction of the physical states can be accomplished explicitly. But the decoupling of ghosts is more general.

In addition to being gauge invariant, $N(A)$ is also scale invariant. For a solution to $\partial^\mu A^U_\mu=0$ for example there is another solution on another gauge orbit where $A(x)\rightarrow \lambda A(\lambda x)$ and $U(x)\rightarrow U(\lambda x)$. Then the two gauge orbits are related by a scale transformation and the Gribov copies on the two orbits are in one-to-one correspondence. Thus gauge inequivalent configurations related by a scale transformation have the same $N(A)$.

An explicit counting of copies is made possible by a restriction to configurations of finite norm in Euclidean space,
\begin{equation}
||A||\equiv\int d^4x A^a_\mu(x) A^{a\mu}(x)<\infty.
\end{equation}
Every gauge orbit does have configurations of finite norm and so this constitutes a partial gauge fixing. The allowed gauge orbits are now characterized by the value of the norm functional and at the stationary points the remaining gauge fixing $\partial^\mu A_\mu=0$ is satisfied. This makes the number of copies $N(A)$ well defined. The norm along a gauge orbit has reached a global minimum at the point it intersects the fundamental modular region by definition. In other words the gauge fixed hypersurface restricted to finite norm contains within it the FMR. Our focus then is on the $1/(1+N(A))$ factor that is needed to compensate for the copies in this space of finite norm configurations.

In particular we would like to determine how $N(A)$ depends on the amplitudes $A_\mu(k)$ and wave vectors $k$ of the fourier modes in a finite Euclidean volume $V=L^4$,
\begin{equation}
A_\mu(x)=\frac{1}{L^2}\sum_k A_\mu(k)e^{i k x}
.\end{equation}
Note that once the volume is fixed then we can no longer consider scale transformations, since the spatial extent of a finite norm configuration should also change under a scale transformation. In section 3 we study configurations that are approximations to finite volume plane waves and which allow a counting of Gribov copies to be performed. In section 4 we study the distribution of copies along the gauge orbits and discuss the connection to confinement.

\section{Implications}
In this section we shall discuss the implications of two of the results that follow from the counting of Gribov copies, as described in the next section.
\begin{itemize}
\item For $A_\mu(k)A^\mu(-k)/k^2$ less than some constant, $N$ identically vanishes.
\item When $N$ is large it grows approximately as a positive power of $A_\mu(k)A^\mu(-k)/k^2$.
\end{itemize}

The first result is a consequence of the discrete nature of $N$ and it means that there is a region within the space of finite norm configurations for which $N$ vanishes. The gauge fixed hypersurface within this subregion will be a subregion of the fundamental modular region.

Given these properties, we would like to illustrate the effect that the $1/(1+N)$ factor in the functional integral can have on the gauge-fixed propagator.  For this we shall develop a toy model that shows the effect on the propagator of an otherwise free theory. For a free theory the generating functional in momentum space can be written as a product of simple one-dimensional integrals, one for each $k$. We replace $A_\mu(k)$ by $A$ and consider the Euclidean form of the integrals where a space-like momentum is mapped to a positive $k^2$. And further we shall temporarily suppose that the $1/(1+N)$ factor also factorizes over the momentum modes; we shall comment more on this below. We thus consider the toy generating functional based a one-dimensional integral:
\begin{equation}
Z(j)=\int_{-\infty}^{\infty}dA \frac{1}{1+N(\frac{\Lambda^4}{k^2}A^2)}\exp(-\frac{1}{2}k^2A^2+jA)
.\label{e11}\end{equation}
The toy propagator is
\begin{equation}
G(k^2)=\left.\frac{1}{Z(j)}\frac{d^2}{d j^2}Z(j)\right|_{j=0}\quad\stackrel{N=0}{\longrightarrow}\quad\frac{1}{k^2}.
\end{equation}

The dependence of $N$ on $\frac{\Lambda^4}{k^2}A^2$ follows from the properties of $N$ noted above and dimensional analysis. We shall model this dependence as follows:
\begin{equation}
N=\left\{\begin{array}{cc} 0 &A^2<\frac{k^2}{\Lambda^4} \\ \left(\frac{\Lambda^4}{k^2}A^2\right)^{a/2}-1\quad &A^2>\frac{k^2}{\Lambda^4}\end{array}\right.
\label{e10}\end{equation}
The appearance of a physical mass scale $\Lambda$ corresponds to an explicit breaking of scale invariance; it is a physical scale below which Gribov copies are important and above which they are not. From (\ref{e11}) we see that $\Lambda$ determines the cross over between the infrared and ultraviolet regimes as follows: for $k^2\gg \Lambda^2$ the exponential factor constrains the range of $A$ that contributes to the integral, while for $k^2\ll \Lambda^2$ it is the $1/(1+N)$ factor that constrains the range of $A$.

It is interesting that a mass scale $\Lambda$ enters as soon as the classical structure of the gauge configuration space is put into a quantum mechanical context, due to the $A^2/k^2$ dependence of the former versus the $k^2 A^2$ dependence of the latter. It is the classical effects of the gauge field interactions that are responsible for a nontrivial $N$. The direct quantum effects of interactions are being ignored here, but they also give rise to a mass scale through dimensional transformation. It seems reasonable to assume that these two mass scales are similar.

For this model $G(k^2)$ can be determined analytically, but we don't give the messy expression here. The leading behavior at $k^2\ll \Lambda^2$ is
\begin{equation}
G(k^2)=\left\{\begin{array}{cc} \frac{1}{k^2}\frac{1}{\ln(\Lambda^2/k^2)}& a=1 \\ \frac{\sqrt{2\pi}}{4}\frac{1}{\Lambda^2}&a=2\\ \frac{2}{3}\ln(\Lambda^2/k^2)\frac{k^2}{\Lambda^4}\quad &a=3\\ \frac{1}{3}\frac{a-1}{a-3}\frac{k^2}{\Lambda^4}&a>3\end{array}\right\}+...
\end{equation}
Our explicit counting results of the next section show that $a$ is certainly greater than 3 and is closer to 5. The leading behavior at $k^2\gg \Lambda^2$ is
\begin{equation}
G(k^2)=\frac{1}{k^2}\left(1-a\sqrt{\frac{2}{\pi}}\frac{\Lambda^2}{k^2}\exp(-\frac{1}{2}\frac{k^4}{\Lambda^4})+...\right)
\label{e8}\end{equation}
We can also consider a smooth analytic continuation to negative $k^2$, and then we find that $G(k^2)\rightarrow(1-a)/k^2$ for $-k^2\gg\Lambda^2$.

We can compare these results to the original analysis of Gribov \cite{n1,n2}, which described an attempt to remove the $1/(1+N)$ factor by instead restricting the configuration space to the first Gribov region. The implementation of this restriction in his semi-perturbative approach had the effect of introducing into the functional integral another factor instead of $1/(1+N)$. This factor happens to factorize over the fourier modes and in the notation of our toy model takes the form $\exp(-\frac{\Lambda^4}{k^2}\frac{A^2}{2})$. This produces the propagator
\begin{equation}
G(k^2)=\frac{1}{k^2+\frac{\Lambda^4}{k^2}}=\left\{\begin{array}{cc} \frac{k^2}{\Lambda^4}+...& k^2\ll \Lambda^2 \\ \frac{1}{k^2}\left(1-\frac{\Lambda^4}{k^4}+...\right)\quad&k^2\gg \Lambda^2\end{array}\right.
.\label{e9}\end{equation}
This propagator vanishes with $k^2$ in the infrared, which is also the case for our model for $a>3$. Although this infrared suppression of propagation is different from a conventional mass, there are no longer massless excitations in the infrared and in this sense a mass gap has formed. The vanishing of the propagator implies that it violates of positivity \cite{a4}, and in the context of gluon propagation in QCD this is expected for confinement. Violation of positivity is also found in lattice studies \cite{a15}.

The main difference in the two models is in the ultraviolet behavior, where our propagator approaches the free propagator exponentially quickly according to (\ref{e8}), in contrast to the power law approach in (\ref{e9}). A power law approach to perturbative behavior is common to all versions of the Gribov-Zwanziger framework \cite{n1,n14,a4,n12,a21}.

An exponential approach to the free propagator has a significant implication. The operator product expansion for QCD with massless quarks relates the asymptotic behavior of the gluon propagator to various condensates.
\begin{equation}
(k^2 G(k^2))^{-1}\stackrel{k^2\rightarrow\infty}{\sim}a_1+a_2\frac{\langle G_{\mu\nu}G^{\mu\nu}\rangle}{k^4} +a_3\frac{\langle\overline{q}q\rangle^2}{k^6}+...
\label{e6}\end{equation}
Thus if a gluon propagator has ultraviolet behavior as in (\ref{e8}) or (\ref{e9}) it is associated with a vanishing or non-vanishing gluon condensate respectively. The gluon condensate in QCD is in turn related by the trace anomaly to $\frac{1}{4}\langle T_\mu^\mu\rangle$, which by Lorentz invariance is the vacuum energy. Our model yields vanishing vacuum energy, and at a superficial level this could be said to reflect the classically scale invariant nature of Gribov copies.

We have traced how the quantity $\Lambda$ that reflects the breaking of scale invariance enters the description of  Gribov copies. One might have expected the breakdown of scale invariance to be manifested by a non-vanishing gluon condensate of order $\Lambda^4$. But we see that this is not the case. In fact we can relate the absence of a gluon condensate to the discrete nature of $N$; because $N$ is discrete it can and does remain identically zero until $\frac{\Lambda^4}{k^2}A^2$ is sufficiently large as in (\ref{e10}). This is the cause of the exponentially small effects of Gribov copies as in (\ref{e8}) in the regime $k^2\gg \Lambda^2$ since the contributions from $A^2\gg1/k^2$ are exponentially suppressed. The effects of Gribov copies are thus not sufficient to produce a local gluon condensate, which as the operator product expansion demonstrates is a modification of the short distance behavior of the theory. This reflects the nonlocal and long distance nature of Gribov copies.

Finally we comment further on the likely false assumption that $1/(1+N)$ factorizes in the space of fourier modes, which led to our toy model. For the toy model the region over which $N$ identically vanishes is a hypercube in the space of all $A_\mu(k)$. In a more realistic description the hypercube is probably replaced by some other shape. But it clear that the reasoning of the previous paragraph depends on the mere existence of a region where $N$ identically vanishes, and not on its shape. Note also that in QCD the coupling falls in the ultraviolet, and thus the effect of Gribov copies in the ultraviolet is even less than what our toy model indicates.

\section{Counting}
For our purposes here we shall consider static gauge field configurations, in which case the Landau and Coulomb gauge conditions become identical. We restrict further to static spherically symmetric configurations. As we shall see, these configurations have finite $N$ when restricted to finite spatial norm. We take the gauge group to be $SU(2)$ with gauge field $A_\mu(x)=A^a_\mu(x)\sigma^a/2i$. Gauge configurations of the form
\begin{equation}
A^a_i(x)=\frac{f(r)}{r}\varepsilon_{iab}\frac{x^b}{r}
\label{e2}\end{equation}
automatically satisfy the gauge condition $\partial_i A_i^a=0$. By an appropriate choice of $f(r)$ we shall be able to approximate plane waves having only spatial oscillations, and thus model the spacelike $k^2$ of the previous section.

The following gauge transformations produce more general spherically symmetric gauge configurations,
\begin{equation}
U(x)=\cos(\alpha(r)/2)\mathbf{1}+i\sin(\alpha(r)/2)\frac{x^a}{r}\sigma^a
.\label{e3}\end{equation}
For the transformed gauge field to be a Gribov copy it must satisfy $\partial_i(A^U)_i^a=0$, and this leads to the Gribov pendulum equation \cite{n1,n2,n7}:
\begin{equation}
r^2\alpha''(r)+2r\alpha'(r)-2(gf(r)+1)\sin(\alpha(r))=0.
\label{e4}\end{equation}
The classical scale invariance corresponds to solutions related by $f(r)\rightarrow f(\lambda r)$ and $\alpha(r)\rightarrow\alpha(\lambda r)$. Here we only study the classical configuration space and thus we treat the coupling $g$ as a constant; also the mass scale $\Lambda$ will not appear in this section or the next.

From (\ref{e2}), $f(r)\rightarrow0$ as $r\rightarrow0$ at least as fast as $r$ to have a regular configuration. To maintain regularity under the gauge transformation, $\alpha(r)\rightarrow 2n\pi$ for integer $n$ as $r\rightarrow0$ so that $U\rightarrow\pm\mathbf{1}$. For a given $\alpha(0)$ the set of solutions of the pendulum equation (the set of copies) is parameterized by the ``initial condition'' $\alpha'(0)$. Since the equation is invariant under both  the constant shift $\alpha(r)=\alpha(r)+2n\pi$ and the reflection $\alpha(r)\rightarrow-\alpha(r)$, and since the $g$ dependence can be absorbed into $f(r)$, without lack of generality we may set $\alpha(0)=0$, $\alpha'(0)>0$ and $g=1$.

The infinite number of copies parameterized by $\alpha'(0)$ reduce down to a finite number, possibly zero, under the further restriction to finite norm configurations. For the configuration in (\ref{e2}) the requirement is
\begin{equation}
||A||\equiv\int d^3x A^a_i(x) A^a_i(x)=8\pi\int_0^\infty dr f(r)^2<\infty.
\end{equation}
For this to be also true of $||A^U||$, the gauge transformation must have the property that $\alpha(r)\rightarrow 2n\pi$ as $r\rightarrow\infty$ for some integer $n$. Since $\alpha(0)=0$ this implies that the gauge transformation has a winding number $n$. The pendulum equation is such that these finite norm solutions only exist for a finite set of $\alpha'(0)$. On the other hand the infinite set of infinite norm solutions are characterized by $\alpha(r)\rightarrow (2n+1)\pi$ as $r\rightarrow\infty$ \cite{n2}.

Since we work in the space of finite norm configurations, by Gribov copies we shall henceforth refer only to the finite norm solutions. Because (\ref{e4}) can be efficiently integrated as an initial value problem, copies can be counted by simply observing the behavior of $\alpha(\infty)$ as a function of $\alpha'(0)$.\footnote{Numerically we integrate $\alpha(r)$ out to $r=10^{10}$.} For example for two values $\alpha'(0)$, $\alpha(\infty)$ may jump from some value $(2n_1+1)\pi$ to another value $(2n_2+1)\pi$. We then know by continuity that there are at least $|n_1-n_2|$ copies between these two values of $\alpha'(0)$. The numerical task then is to calculate a set of $\alpha(\infty)$ for a sufficiently fine grained set of $\alpha'(0)$, and then obtain a value for $N$ by adding the $|n_1-n_2|$ for all the jumps in $\alpha(\infty)$. Copies can be missed when $\alpha(\infty)$ jumps forward and backward within the $\alpha'(0)$ stepsize.  As $N$ becomes larger the copies are becoming closer together, and thus the stepsize of $\alpha'(0)$ must be appropriately small to approach the true $N$. In addition the maximum value of $\alpha'(0)$ beyond which there are no further copies is increasing for larger $N$.

We shall use these spherical configurations to approximate spatially dependent plane waves. We thus consider configurations that oscillate as a function of $r$ and which are modulated by an amplitude which vanishes in the small and large $r$ limits. When the radius of the amplitude peak is much larger than the wavelength of the oscillation we are approaching plane waves. For the radial profile we choose the form
\begin{equation}
f(r)=A e^{r/R}(1-e^{r/R})^\rho\sin(kr),\quad \rho=1\mbox{ or }2, 
\end{equation}
where we need to choose $R$ large compared to $1/k$. $k$ and $A$ are analogous to the $k$ and $A$ of the toy model in the previous section. Holding $R$ and $\rho$ fixed is analogous to holding the volume fixed, while a scale transformation $f(r)\rightarrow f(\lambda r)$ is equivalent to $k\rightarrow \lambda k$, $A\rightarrow A$, and $R\rightarrow R/\lambda$.
\begin{center}\includegraphics[scale=0.5]{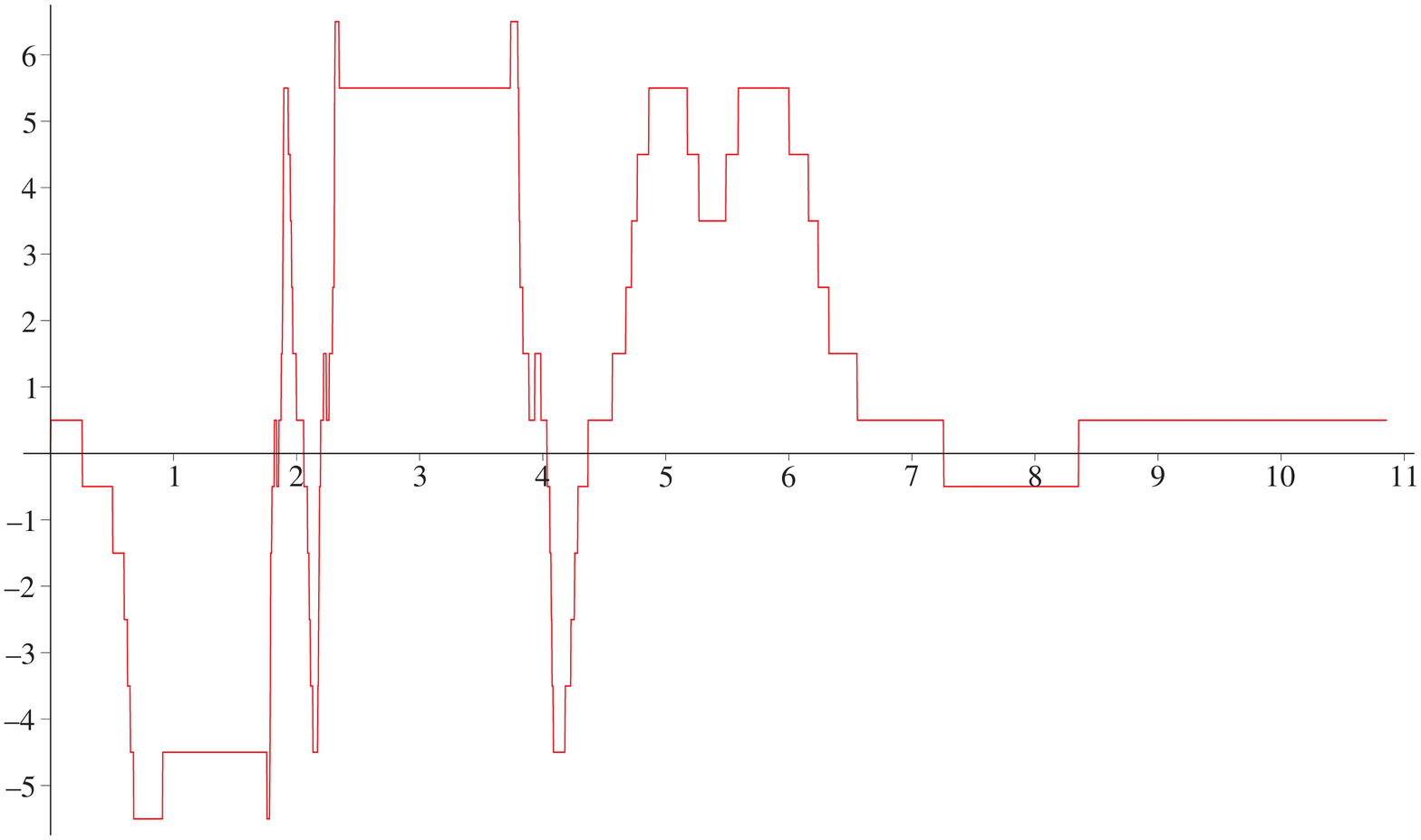}\end{center}
\vspace{-1ex}\noindent Figure 1: $\alpha(\infty)/2\pi$ versus ${\rm arcsinh}(\alpha'(0))$ for $A=100$, $k=1$, $R=30$, $\rho=1$.
\vspace{2ex}

We choose the discretization $\alpha'(0)_n=\sinh(n\Delta)$ for integers $n$ and small $\Delta$, since then the copies are spread out quite evenly as a function of $n\Delta$ (or ${\rm arcsinh}(\alpha'(0))$).  In Fig.~(1) we present an example of the behavior of $\alpha(\infty)/2\pi$ for $A=100$, $k=1$, $R=30$ and $\rho=1$, where we obtain $N=80$. Copies (of finite norm) exist where $\alpha(\infty)/2\pi$ crosses integer values. These values are the winding numbers of the gauge transformations, and we see in this example that there are 10 copies of zero winding number.

Our counting results show that $N$ is zero unless $A/(k+c_1)>c_2$, where the constants  depend on our modeling of plane waves. We plot values of $A$ versus $k$ at which the first copies occur in Fig.~(2) for $(R,\rho)=(30,1),(100,1),(30,2)$. The linear relationship holds very well and we find that $c_1\approx0.45, 0.18, 0.0$ respectively. We understand the offset $c_1$ as follows. For small $k$ we are not approaching a constant gauge field, but rather a spherical configuration with a characteristic size $R$, corresponding to a superposition of plane wave modes. $c_1$ tends to zero when the radial profile is more strongly suppressed near $r=0$, as when $\rho=2$.
\begin{center}\includegraphics[scale=0.5]{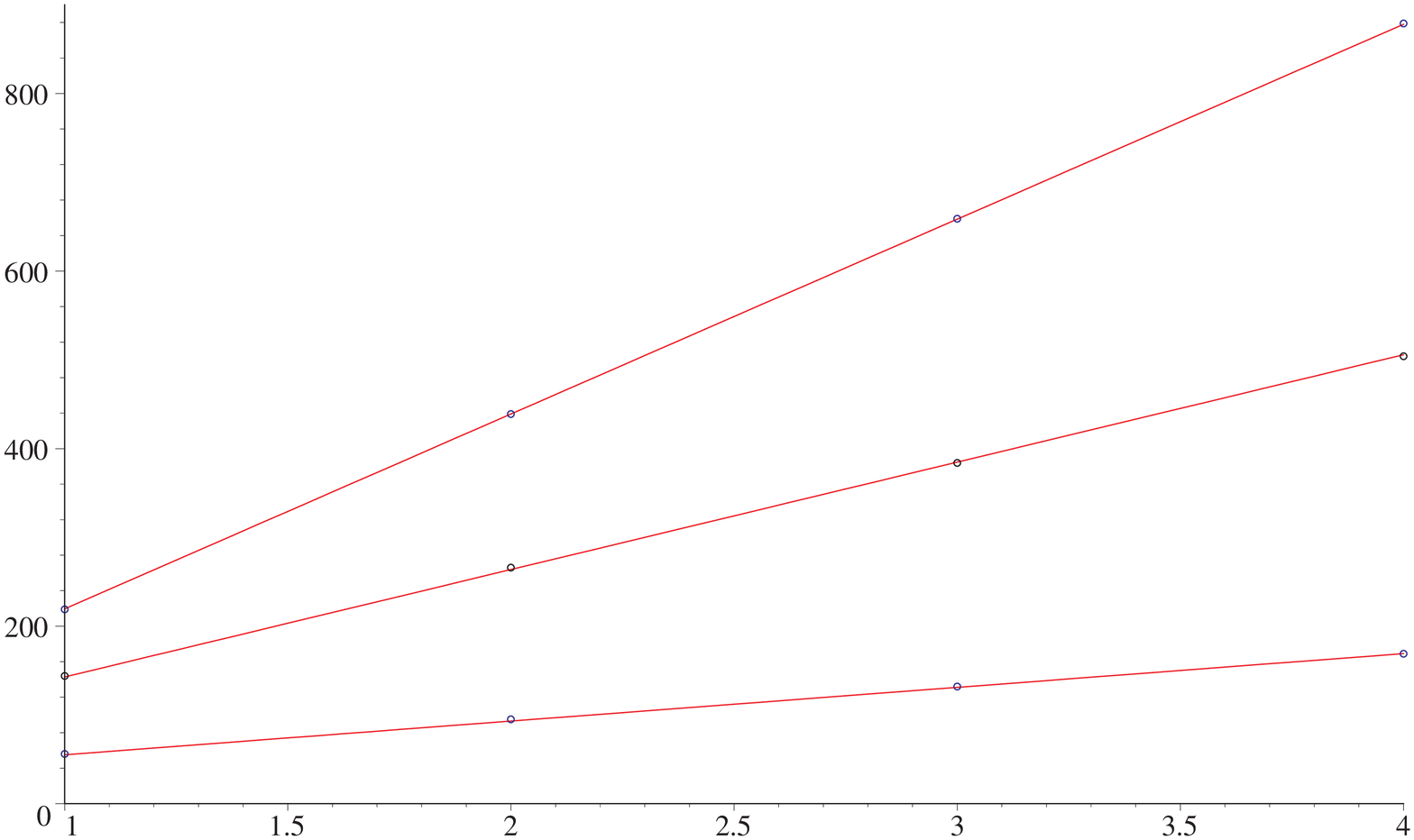}\end{center}
\vspace{-1ex}\noindent Figure 2: $A$ versus $k$ at which the first copies appear for different models of plane waves. $(R,\rho)=(30,2),(100,1),(30,1)$ from top to bottom.
\vspace{2ex}

As motivated by the last section, the main point of this section is to find the behavior of $N$ when it is large. To investigate this we study the dependence of $N$ on $A$ for large $N$ and fixed $k$. In Fig.~(3) we display $\log(N)$ versus $\log(A)$ for the cases $(k,R,\rho)=(2,30,1), (1,100,1), (1, 30,2)$. The number of copies we have calculated per configuration ranges up to nearly 14000. To achieve a single value of $N$ this large we generate over 350000 numerical solutions of the Gribov pendulum equation.\footnote{We used the rk8pd method of the GSL C package.} The roughly linear behavior seen in Fig.~(3) indicates a power law growth $N\propto A^a$ with $a\approx4.9, 4.9, 5.5$ respectively. This power law behavior leads to the model we investigated in the previous section.
\begin{center}\includegraphics[scale=0.5]{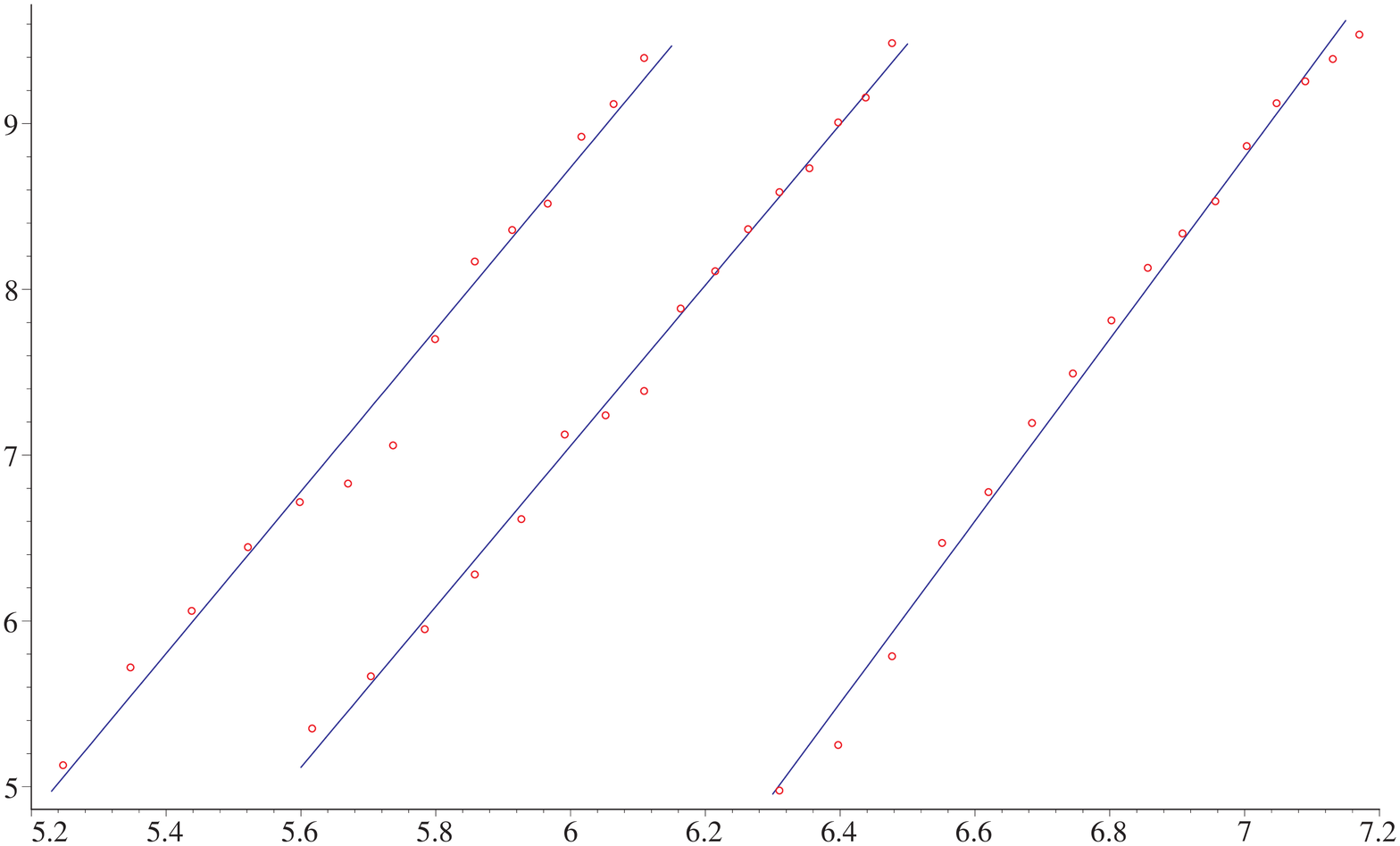}\end{center}
\vspace{-1ex}\noindent Figure 3: $\log(N)$ versus $\log(A)$. $(k,R,\rho)=$ (2,30,1), (1,100,1), (1, 30,2) from left to right.
\vspace{2ex}

\section{Towards confinement}
Since the Gribov pendulum equation has a symmetry $\alpha(r)\rightarrow-\alpha(r)$, every copy is duplicated under $\alpha'(0)\rightarrow-\alpha'(0)$, and thus the total number of copies is even.\footnote{The $N$'s given in the previous section ignored this duplication.} Let us continue to focus on the copies with $\alpha'(0)>0$. Consider the values of $\alpha(\infty)$ for $\alpha'(0)\rightarrow 0+$ and $\alpha'(0)\rightarrow+\infty$ respectively and let us refer to these two values as $\alpha_0(\infty)$ and $\alpha_\infty(\infty)$. For example in Fig.~(1) both values are $+\pi$. From our numerical analysis we find that $\alpha_\infty(\infty)$ is always $+\pi$ while $\alpha_0(\infty)$ can be $\pm\pi$. When there are no copies $\alpha_0(\infty)=\pi$. Often there is a range of $A$ for which only one copy exists (along with its duplicate), and this copy first appears (as $A$ increases) when $\alpha_0(\infty)$ changes from $\pi$ to $-\pi$. The copy has a value of $\alpha'(0)$ which is infinitesimally close to zero for some value of $A$.

In this case the field configuration is related to either of its close copies by an infinitesimal gauge transformation $U(x)=1+X(x)$ and this implies a vanishing eigenvalue of the FP operator, ${\cal D}_A=-\partial_i D_i$,
\begin{equation}
\partial_i A_i=0=\partial_i(U^{-1}A_i U+\frac{1}{g}U^{-1}\partial_i U)=-\frac{1}{g}{\cal D}_AX(x)
.\label{e17}\end{equation}
In such a case the curvature of the norm functional is also vanishing,
\begin{eqnarray}
||A^U||&=&||A||+2\int d^3x\mbox{tr}(X\partial_i A_i)+\int d^3x\mbox{tr}(X^\dagger{\cal D}_AX)+{\cal O}(X^3)\label{e5}\\&=&||A||+{\cal O}(X^3).\nonumber
\end{eqnarray}

Thus as $A$ increases the first Gribov boundary is being crossed and an originally positive eigenvalue ${\cal D}_A$ is turning negative. An analogy is to the function $(x^2+a^2)^2$ when $a^2$ changes from positive to negative; the curvature at $x=0$ turns negative and two new minima appear. Thus we expect the original configuration to pick up a negative eigenvalue while the two copies do not, and so the copies remain in the first Gribov region while the original configuration moves outside. The two copies also remain in the FMR, which thus incorporates a reflection symmetry, and it can be checked numerically that the copies have smaller norm as expected. The crossing is occurring where the FMR and first Gribov boundaries coincide. This bifurcation picture is described in \cite{a14}.

As $A$ increases further Gribov copies can continue to appear infinitesimally close to the original configuration, and thus more Gribov boundaries are being crossed and ${\cal D}_A$ gains more negative eigenvalues. Thus increasing $A$ is pushing the configuration into ever higher Gribov regions. For the case $R=30$, $\rho=2$, $k=1$ we plot $\alpha_0(\infty)$ versus $A$ in Fig.~(4), which thus shows the values of $A$ at which Gribov horizons are being crossed.
\begin{center}\includegraphics[scale=0.35]{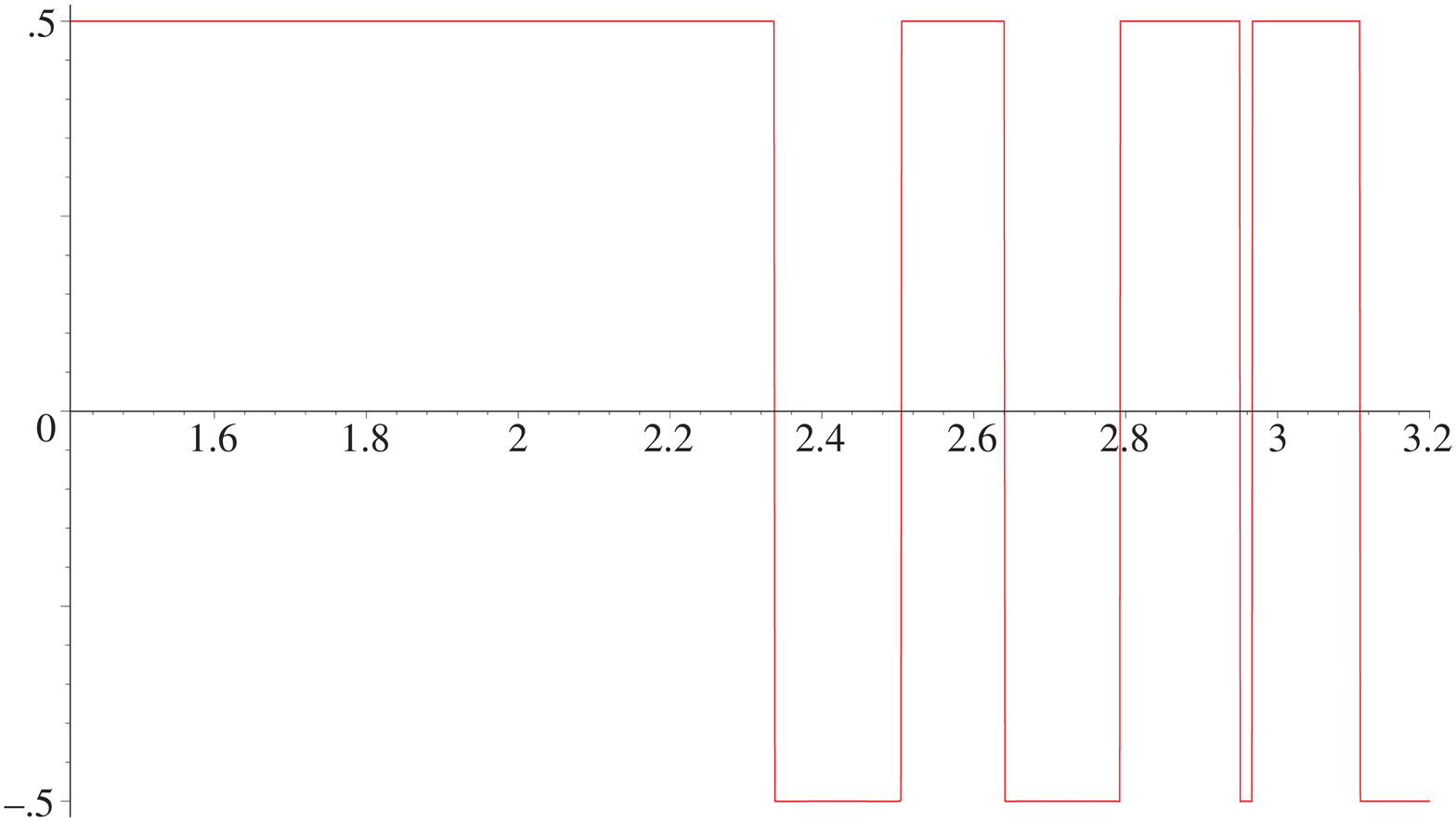}\end{center}
\vspace{-1ex}\noindent Figure 4: $\alpha(\infty)/2\pi$ versus $\log_{10}(A)$ for $\alpha'(0)\rightarrow 0+$.
\vspace{2ex}

The gauge orbit can cross a Gribov boundary anywhere all along the orbit, and this corresponds to a pair of copies appearing at a finite $\alpha'(0)$ (along with a duplicate pair at $-\alpha'(0)$). The difference in $\alpha'(0)$ for the pair starts from zero and increases as $A$ is increased. The analog now is the function $x(x^2+a^2)$ as $a^2$ turns negative, so that a new minimum and maximum appear. This is how most copies appear and in fact it sometimes happens that the first copies appear in this way rather than as described above \cite{a14}. In this case the original configuration stays within the FMR and the new pair is near a Gribov boundary that does not coincide with the FMR boundary. The new pairs may also be related to the original configuration by gauge transformations of nonzero winding number.

By the time there are many copies then it will often happen that two copies are separated by a small but finite gauge transformation. Then the $X(x)$ appearing in (\ref{e5}) is determined by two solutions $\alpha_1(r)$ and $\alpha_2(r)$ with $\alpha'_1(0)\approx\alpha'_2(0)$ and $\alpha_1(\infty)=\alpha_2(\infty)=2\pi n$. The gauge transformation determined by $\alpha(r)=(1-s)\alpha_1(r)+s\alpha_2(r)$ with $s$ changing from 0 to 1 moves along a gauge orbit connecting the two Gribov copies. This path in configuration space moves off the gauge fixed hypersurface but it remains within the space of finite norm. Thus while $X(x)$ will no longer represent a zero mode of ${\cal D}_A$, if it is small it can still represent a direction at the stationary point of the norm functional for which the curvature $C\equiv\int d^3x\mbox{ tr}(X^\dagger{\cal D}_AX)/\int d^3x\mbox{ tr}(X^\dagger X)$ is small, $|C|\ll1$. The two nearby copies typically have curvatures of opposite sign corresponding to being on either side of a Gribov horizon; an eigenvalue of ${\cal D}_A$ is changing sign as the Gribov horizon is crossed. The difference $\Delta\alpha'(0)$ is a measure of how far apart the two copies are. We expect a linear relationship $|C|\propto\Delta\alpha'(0)$ for small separations. To illustrate, the representative function $x(x^2-3a^2)/6$ has two stationary points at $x=\pm a$ with respective curvatures also equal to $\pm a$.
\begin{center}\includegraphics[scale=0.5]{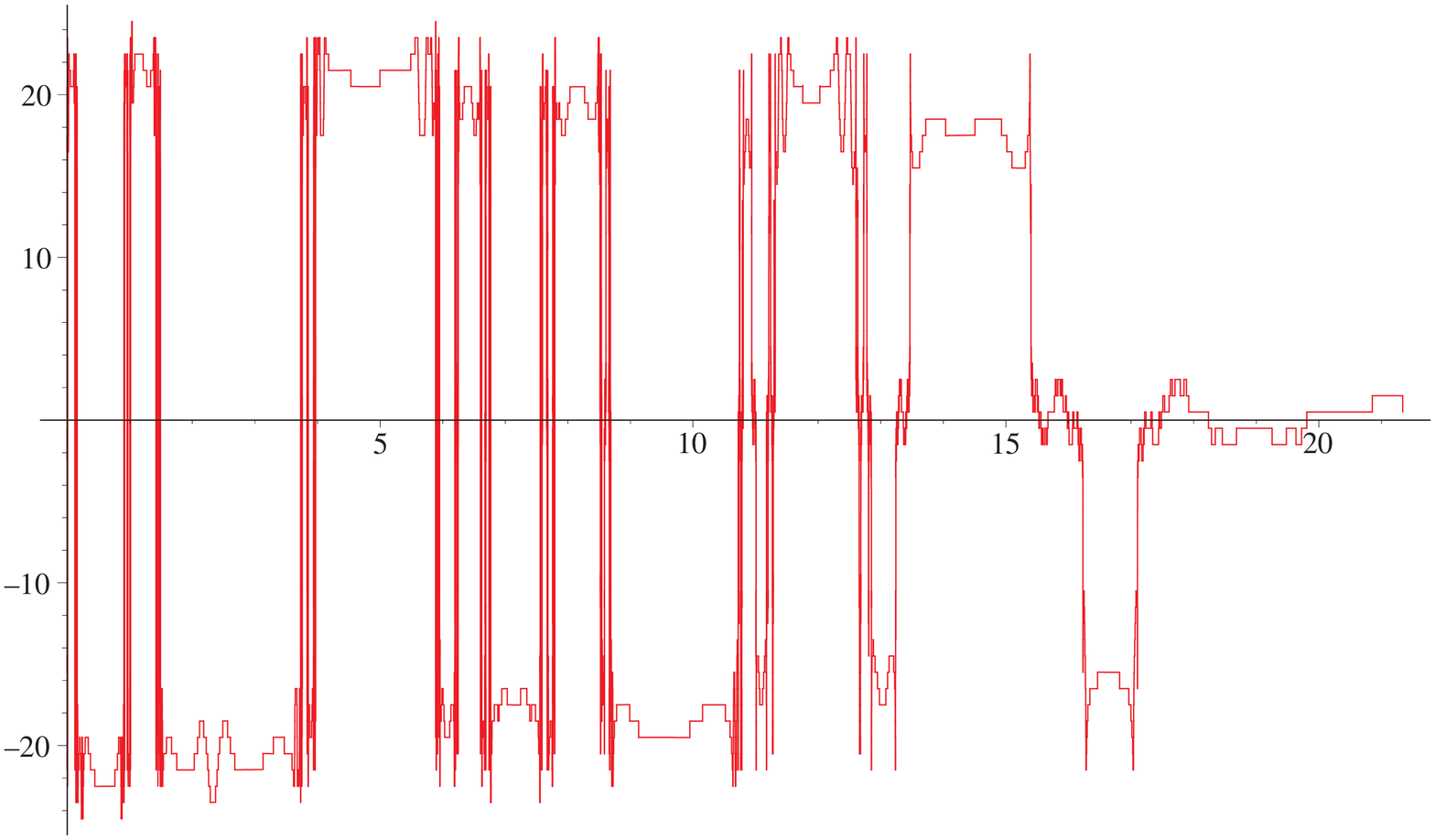}\end{center}
\vspace{-1ex}\noindent Figure 5: $\alpha(\infty)/2\pi$ versus ${\rm arcsinh}(\alpha'(0))$ for a large $N$ case.
\vspace{2ex}

Thus if we obtain the distribution of $\Delta\alpha'(0)$ values for the ``close'' pairs of copies, on a gauge orbit with many copies, we are obtaining a distribution of the small curvatures, which are the small diagonal matrix elements of ${\cal D}_A$. We expect these matrix elements to approach the eigenvalues in the small curvature limit, and thus we are also obtaining a distribution of the magnitudes of the small eigenvalues, the $|\lambda_i|$'s. This distribution enters the functional integral via the $|\det({\cal D}_A))|$ factor in (\ref{e1}), which is also being sampled at all the gauge copies along the orbit.

We need only consider pairs of copies having the same winding number, since only then can the relative gauge transformation become small. We consider two ways to construct the distribution. In the first method we consider the separations of all the nearest neighbor pairs of copies. In the second method, for each copy we take the separation to the neighboring copy which is the closest. This may count the same separation twice, but this corresponds to trying to find the smallest curvature at each copy.

We give in Fig.~(5) an example of $\alpha(\infty)/2\pi$ versus ${\rm arcsinh}(\alpha'(0))$ (where the latter is evaluated for discrete values $n\Delta$) for $A=1300$, $k=1$, $R=30$, $\rho=2$, $\Delta=1/14000$. The resulting estimate for $N$ is 13856 copies (ignoring the duplicates). Each separation is some multiple of $\Delta$. To study the distribution in these separations we plot the log of the number that each separation occurs versus the log of the separation, for each of the two methods. The results in Fig.~(6) and (7) have lines added for comparison with slopes  of $-1$ and $-4/3$ respectively. The first method suggests that the density of eigenvalues scales as $|\lambda|^{-1}$. We will consider this case in the following, although the second method could indicate that there is more infrared enhancement. We compare this density to the density of eigenvalues of the free FP operator $(-\nabla^2)$ which scales as $\lambda^{-1/2}$.\footnote{In the free case $dk=d\lambda/(2\sqrt{\lambda})$ and the eigenvalues are evenly spaced in $k$ for planes waves of fixed direction, which is what we are modeling.} This suggests that the density of small eigenvalues of ${\cal D}_A$ is enhanced in the infrared, in relation to the free case, by a factor $\approx|\lambda|^{-1/2}$.
\begin{center}\includegraphics[scale=0.5]{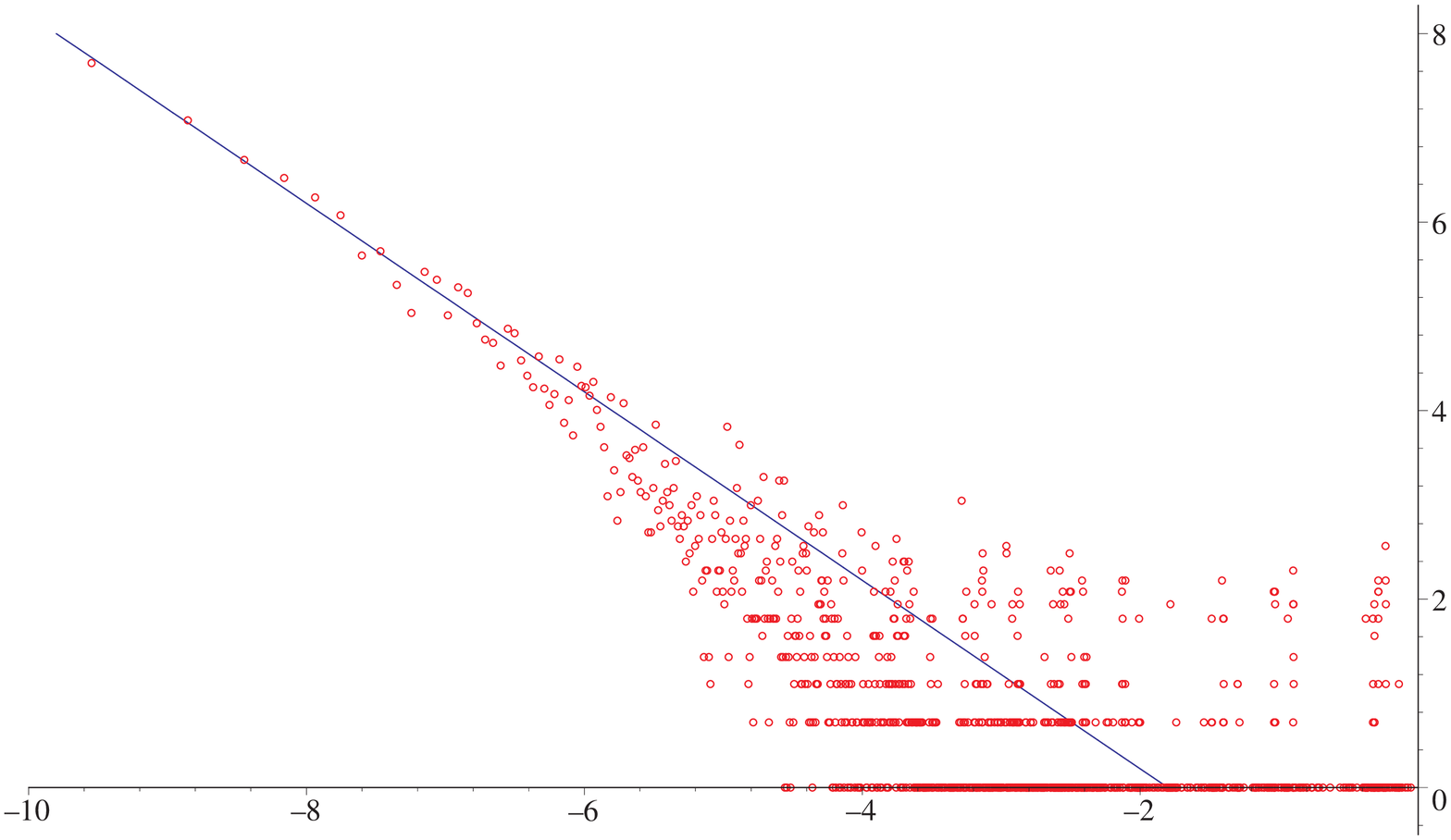}\end{center}
\vspace{-1ex}\noindent Figure 6: (log of the frequency of each separation) versus (log of the separation) by method 1. The line has slope $-1$.
\vspace{2ex}
\begin{center}\includegraphics[scale=0.5]{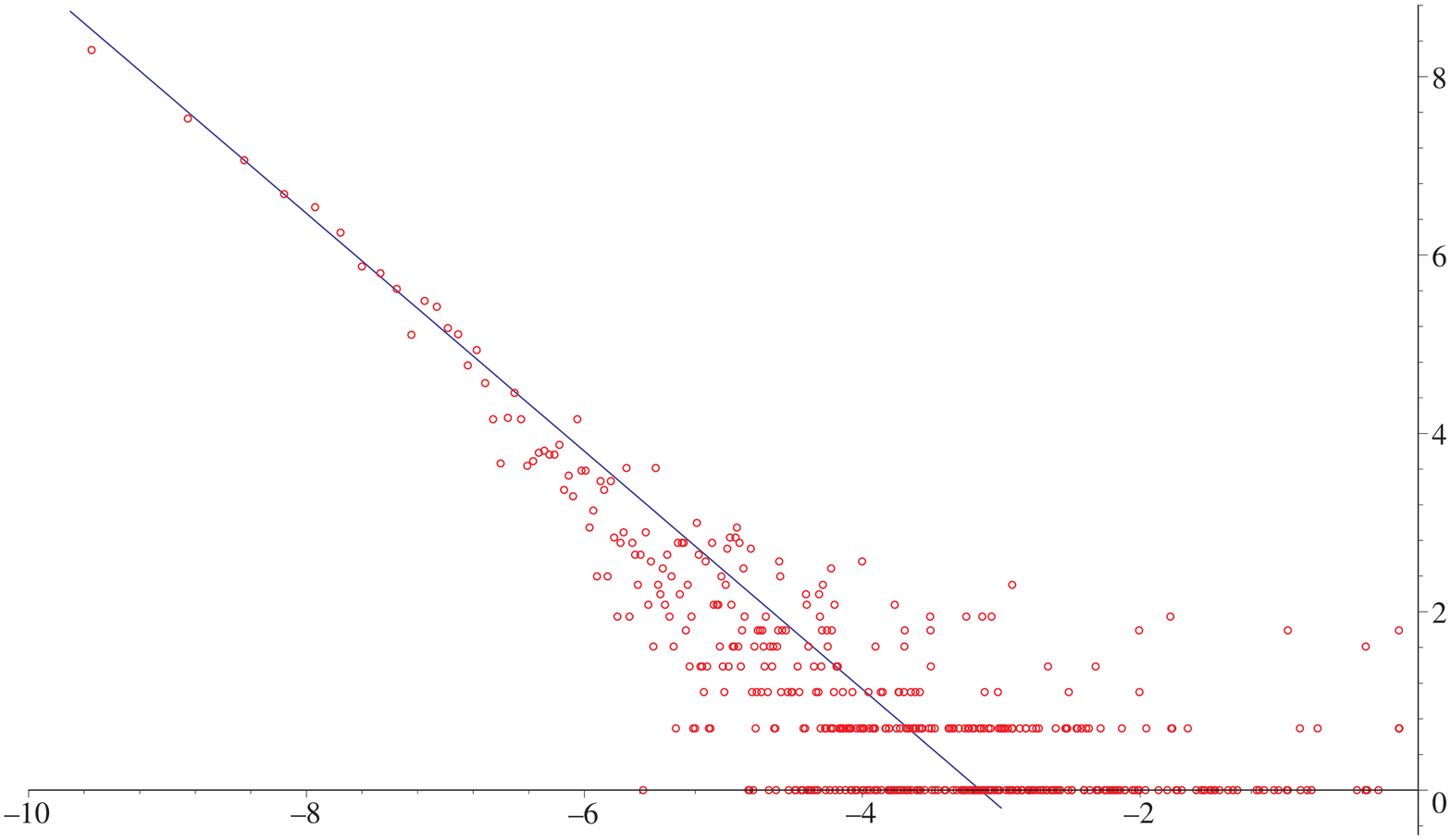}\end{center}
\vspace{-1ex}\noindent Figure 7: (log of the frequency of each separation) versus (log of the separation) by method 2. The line of slope $-4/3$.
\vspace{2ex}

For a non-abelian gauge theory, the Coulomb gauge Fadeev-Popov operator ${\cal D}_A=-\nabla\cdot D$ determines the classical instantaneous Coulomb potential \cite{n4},
\begin{equation}
A_0^a(x)=g\int d^3y[{\cal D}_A^{-1}(-\nabla^2){\cal D}_A^{-1}]_{xy}^{ab}\sigma^b(y)
.\end{equation}
The color charge can have quark and gauge field contributions, $\sigma^b=\sigma_{\rm qu}^b-f^{bcd}A^c\cdot E^d$, with $\nabla\cdot A^a=\nabla\cdot E^a=0$.

Then the Coulomb self-energy of a static point color charge in a finite volume is proportional to \cite{n5}
\begin{equation}
[{\cal D}_A^{-1}(-\nabla^2){\cal D}_A^{-1}]_{xx}^{aa}\propto\sum_n \frac{F_n}{\lambda_n^2},
\end{equation}
where $F_n=\langle\lambda_n|(-\nabla^2)|\lambda_n\rangle$ is a diagonal matrix element in the Faddeev-Popov eigenstates. In the large volume limit with an ensemble average over the gauge field we can write the self-energy as \cite{n5}
\begin{equation}
{\cal E}=\int_0^{\lambda_{max}}d\lambda\rho(\lambda)\frac{F(\lambda)}{\lambda^2}.
\label{e12}\end{equation}
We view this as an integral over $|\lambda|$ and where $\rho(|\lambda|)$ is the appropriately normalized density of eigenvalues. There is the usual ultraviolet divergence cutoff by $\lambda_{max}$, but of interest here is a possible infrared divergence which is related to the growth of the Coulomb potential at large distance. This infrared divergence has been argued to be a necessary condition for confinement \cite{n5}. In the free field limit where the eigenstates are plane waves in 3 dimensions, then $\lambda=\mathbf{k}\cdot \mathbf{k}$, $\rho(\lambda)\propto\lambda^{1/2}$ (from $d^3k\propto d\lambda\sqrt{\lambda}$), $F(\lambda)=\lambda$, and there is no such divergence.

Our previous results indicate that the density of eigenvalues are skewed towards the infrared by a power law enhancement, implying that the eigenvalue density in (\ref{e12}) is enhanced at small $\lambda$, $\rho(\lambda)\rightarrow\rho(\lambda)\lambda^{-1/2}\rightarrow\mbox{ constant}$. We can also expect an infrared enhancement of $F(\lambda)$.  The eigenstates associated with the small eigenvalues are functions that are close to the small gauge transformations $X(x)$, and we find that these latter functions are typically quite rapidly varying with significant overlap with the $f(r)$ profile. Thus the eigenstates typically contain Fourier modes with $\mathbf{k}\cdot \mathbf{k}\gg\lambda$, and this will enhance $\langle\lambda|(-\nabla^2)|\lambda\rangle$ relative to $\lambda$.

 We obtain a simple model of $F(\lambda)$ as follows. Due to the enhanced degeneracy of the eigenvalues of ${\cal D}_A$ relative to $(-\nabla^2)$ at small $\lambda$, the eigenvectors with eigenvalues of $(-\nabla^2)$ in some range $d(k^2)$ correspond to eigenvalues of ${\cal D}_A$ in a smaller range $d\lambda$, where $d(k^2)\propto(d\lambda)\lambda^{-1/2}$. Then we have $\langle k|(-\nabla^2)|k\rangle=k^2=\int_0^{k^2}d(k'^2)\propto\int_0^\lambda(d\lambda')\lambda'^{-1/2}$, and thus $F(\lambda)\propto\lambda^{1/2}$. This an infrared enhancement of $\lambda^{-1/2}$ relative to the free field behavior $F(\lambda)=\lambda$.
 
The enhancement of both $\rho(\lambda)$ and $F(\lambda)$ by $\lambda^{-1/2}$ produces a total infrared enhancement factor of $1/\lambda$ in the integrand of the self energy (\ref{e6}), and this is more than sufficient to produce an infrared divergence. An enhancement factor of $1/\lambda$ implies changing from a Coulomb potential ($1/k^2$ behavior) to a confining linear potential ($1/k^4$ behavior).

\section{Beyond Coulomb gauge}
In Coulomb gauge the gauge fixing on one 3-dimensional time slice is independent of the gauge fixing on another time slice and so results in this gauge apply equally well when the time direction has finite extent, i.e.~at finite temperature. The infinite Coulomb self energy thus persists at high temperature, even above the expected de-confining phase transition, and this is one way to see that the infinite Coulomb self-energy is a necessary but not sufficient condition for confinement \cite{n9}. This then leads us to consider what happens for a covariant gauge fixing such as Landau gauge.

Our counting of Gribov copies of static field configurations applies equally to both Landau and Coulomb gauge conditions. We also recall that ${\cal D}_A^{-1}$ in Landau gauge is the ghost propagator in the gauge field background. Thus the quantity $[{\cal D}_A^{-1}(-\nabla^2){\cal D}_A^{-1}]_{xx}$ that we considered in Coulomb gauge is the 3-dimensional analog of the 4-dimensional ghost loop contribution to a 2-point function, where each vertex corresponds to the ghost current $j^{gh}_\mu=\overline{c}^a\partial_\mu c^a$. The 2-point function in momentum space is
\begin{equation}
\Pi^{\mu\nu}(q^2)=-\int\frac{d^4p}{(2\pi)^4}{\cal D}_A^{-1}(p){\cal D}_A^{-1}(p+q)p^\mu(p+q)^\nu
,\end{equation}
where the color indices are contracted as before. The infrared pile-up of eigenvalues of ${\cal D}_A$ that we have seen in Coulomb gauge must translate into the infrared enhancement of the ghost propagator in Landau gauge. With an implied ensemble average over the gauge field the ghost propagator should take the form
\begin{equation}
{\cal D}_A^{-1}(p)\rightarrow-\frac{i\Lambda^{2\kappa}}{(-p^2-i\epsilon)^{1+\kappa}}\quad\mbox{for }\quad p^2\ll\Lambda^2.
\end{equation}
Here we have reintroduced the scale $\Lambda$. $\kappa=1/2$ is a special value since $\kappa\geq1/2$ leads to an infrared divergence in $\Pi^{\mu\nu}(q^2)$ as $q^2\rightarrow0$. In three space dimensions the enhancement corresponding to $\kappa=1/2$ gave a confining linear potential and was more than sufficient to produce the infrared divergence. But in four space dimensions we would need $\kappa\geq1/2$ to produce an infrared divergence in the Coulomb self-energy.

In section 2 we discussed how the rapidly growing numbers of Gribov copies can cause gluon propagation to be highly damped in the infrared, such as $G(p^2)\approx -p^2/\Lambda^2$ for $-p^2\ll \Lambda^2$. In this case gluonic effects can be integrated out on scales of order $\Lambda$, giving rise to effective nonrenormalizable interactions at this scale. As discussed in \cite{a1} this disrupts a naive power counting, where it is assumed that momenta much smaller than $\Lambda$ can dominate all loops in the infrared. Due to the suppressed gluon propagator, loop momenta of order $\Lambda$ can dominate some loops. These effects can induce effective couplings of the ghosts to other colored fields such as $j^{{\rm gh}\mu}j_\mu/\Lambda_1^2$ where $j^{{\rm gh}\mu}=\overline{c}^a\partial_\mu c^a$ and $j^\mu=\overline{q}^a\gamma_\mu q^a$. As well there are self interactions among the ghosts, such as the interaction $j^{{\rm gh}\mu}j^{\rm gh}_\mu/\Lambda_2^2$ with $\Lambda_1\approx\Lambda_2\approx\Lambda$. (There are also interactions among the color octet versions of the currents.) This has the consequence that ghost loops, each described by $\Pi^{\mu\nu}(q^2)$, may be inserted into diagrams, with each insertion contributing an enhancement factor of order $(\Lambda^2/(-q^2))^{2\kappa-1}$ for $\kappa>1/2$ \cite{a1}.\footnote{This reference also discusses how ghost loop effects can feed back and affect the gluon propagator.}  Due to the ghost coupling to the quark current, this can give rise to a power-law enhanced (or log enhanced for $\kappa=1/2$) anti-screening mechanism. For instance the Wilson loop should be dominated by multiple ghost loop contributions. Perhaps this ghost loop anti-screening mechanism could be developed further as a covariant description of confinement.

In conclusion, we have explored the problem of gauge copies present in the continuum definition of non-abelian gauge theories and found that an explicit counting of these copies leads to insights concerning the infrared behavior of these theories. In the last section we found a pile-up of small eigenvalues of the FP operator. Such an enhancement has been widely discussed in the literature, and in the Gribov-Zwanziger approach it is said to arise due to the constraint on configurations. It is argued that configurations will typically lie close to the Gribov boundary where some eigenvalues of small, due simply to the large dimensionality of the configuration space \cite{n1,n11}. We are instead finding a power-law enhanced density of small eigenvalues by studying the numbers and distributions of copies, in the absence of the constraint on configurations.

Small eigenvalues correspond to small curvatures and thus enhanced gauge fluctuations in the direction perpendicular to the gauged fixed hypersurface. In sections 2 and 3 we described the infrared suppression of the ``physical'' gauge modes within the gauged fixed hypersurface due an effective shrinking of the configuration space in these directions. This is controlled by the rapid growth of $N(A)$. Thus the potentially physical modes are suppressed while unphysical modes, modes that cannot appear in physical states, are enhanced. These two phenomena together present a picture of confinement.  The scale $\Lambda$ characterizes the mass gap of the theory, the scale below which unphysical modes dominate the dynamics and above which the perturbative modes become evident.

In section 2 we also discussed a vanishing contribution to vacuum energy due to the discreteness in the number of Gribov copies. This result is of interest given the present lack of evidence for vacuum energy in QCD with massless quarks \cite{n8}.  We note that the operator product expansion in (\ref{e6}) also shows that a quark condensate does not mimic the effect of a gluon condensate in its effect on the gluon propagator. This is a consistency check on the possibility that the nonperturbative physics responsible for chiral symmetry breaking also does not contribute to vacuum energy in massless QCD. In this respect it would be the same as another possible nonperturbative contribution to vacuum energy, that from instantons, which is known to vanish in massless QCD regardless of chiral symmetry breaking (i.e.~vacuum energy does not depend on the vacuum angle $\theta$ in massless QCD) \cite{a17}.

\section*{Acknowledgments}
This work was supported in part by the National Science and Engineering Research Council of Canada.

\end{document}